\documentclass{PoS}

\usepackage{epsfig}
\usepackage{amsmath}

\newcommand{\bra}{\langle}
\newcommand{\ket}{\rangle}

\newcommand{\be}{\begin{equation}}
\newcommand{\ee}{\end{equation}}

\newcommand{\bea}{\begin{eqnarray}}
\newcommand{\eea}{\end{eqnarray}}
\newcommand{\bean}{\begin{eqnarray*}}
\newcommand{\eean}{\end{eqnarray*}}

\newcommand{\half} {\frac{1}{2}}
\newcommand{\om} {\omega}

\newcommand{\re}{\operatorname{Re}}
\newcommand{\im}{\operatorname{Im}}

\newcommand{\rmR}{{\rm R}}
\newcommand{\rmI}{{\rm I}}

\title{Localised distributions in complex Langevin dynamics}

\ShortTitle{Localised distributions in complex Langevin dynamics}

\author{\speaker{Pietro Giudice}\\
  Universit\"at M\"unster, Institut f\"ur Theoretische Physik,
  M\"unster, Germany\\
  E-mail: \email{p.giudice@uni-muenster.de}}

\author{Gert Aarts\\
  Department of Physics, College of Science, 
  Swansea University, Swansea, United Kingdom\\
  E-mail: \email{g.aarts@swan.ac.uk}}

\author{Erhard Seiler\\
  Max-Planck-Institut f\"ur Physik (Werner-Heisenberg-Institut) 
  M\"unchen, Germany\\
  E-mail: \email{ehs@mppmu.mpg.de}}

\abstract{
Complex Langevin dynamics can be used to perform
numerical simulations of theories with a complex action. In order to 
justify the procedure, it is important to understand the properties 
of the real and positive distribution, which is effectively sampled 
during the stochastic process. In the context of a simple model, we 
study this distribution and relate the results to the recently 
derived criteria for correctness. We demonstrate analytically 
that if the distribution has support only on a strip 
in the complexified configuration space, correct results are expected.
}

\FullConference{31st International Symposium on Lattice Field Theory LATTICE
2013\\
                 July 29 - August 3, 2013\\
                 Mainz, Germany}

\begin{document}

\section{Introduction}

Numerical simulations of lattice field theories with a complex action
are not possible with standard Monte Carlo methods. 
Complex Langevin (CL) dynamics provides an appealing alternative 
since it does not rely on importance sampling, 
see Refs.~\cite{Aarts:2013bla,Aarts:2013uxa} 
for two recent reviews.
In the recent past, the important role  played by the properties of the 
real and positive probability distribution in the complexified 
configuration space, which is effectively sampled during the Langevin process, 
has been clarified \cite{arXiv:0912.3360,arXiv:1101.3270}. 
An important conclusion was that this distribution should be  sufficiently 
localised in order for CL to yield valid results. Moreover, a set of criteria 
for correctness, which have to be satisfied in order for CL to be reliable, 
has been determined.
In this contribution we aim to combine the insights that can be 
distilled from the 
criteria for correctness discussed above with the explicit solution of the 
Fokker-Planck equation (FPE),
adapting the method employed in Ref.~\cite{Duncan:2012tc}. A comprehensive
version of this work has been published~\cite{Aarts:2013qia}.

\section{Complex Langevin dynamics and criteria for correctness}

The model we have studied in this work is described by this simple
partition function:
\be
\label{eq:Z}
Z = \int_{-\infty}^\infty dx\, e^{-S}, 
\quad\quad\quad\quad
 S=\half\sigma x^2+\frac{1}{4}\lambda x^4,
\ee
where the parameters in the action are complex-valued.
This model has been studied shortly after CL was 
introduced~\cite{Klauder:1985ks,Ambjorn:1985iw,Okamoto:1988ru}, 
but no complete solution was given.
We take $\lambda$ real and positive, so that the integral exists, 
while $\sigma$ is taken complex. Analytical results are available: 
a direct evaluation of the integral yields
$Z =  \sqrt{{4\xi}/{\sigma}} \ e^{\xi} \ K_{-\frac{1}{4}}(\xi)$,
where $\xi=\sigma^2/(8\lambda)$ and $K_p(\xi)$ is the modified Bessel 
function of the second kind. Moments $\bra x^n\ket$ can be obtained by 
differentiating with respect to $\sigma$.

We evaluate expectation values numerically, by solving a CL process.
This is done by complexification of the Langevin equation:
\be
\dot z = -\partial_z S(z) +\eta,
\label{eq:lang}
\ee
i.e. we introduce:
\be
z = x+iy, \quad\quad\quad \eta=\eta_\rmR+i\eta_\rmI, \quad\quad\quad \sigma = A+iB.
\ee
The dot in Eq.~(\ref{eq:lang}) denotes differentiating with respect 
to the Langevin time $t$ and the (Gaussian) noise satisfies 
\be
\label{eq:noise}
\bra\eta(t)\eta(t')\ket=2\delta(t-t').
\ee
The normalisation of the real and imaginary noise components is given by
\be
\bra\eta_\rmR(t)\eta_\rmR(t')\ket = 2N_\rmR\delta(t-t'), \quad
\bra\eta_\rmI(t)\eta_\rmI(t')\ket = 2N_\rmI\delta(t-t'), \quad
\bra\eta_\rmR(t)\eta_\rmI(t')\ket =  0,
\ee
with $N_\rmR-N_\rmI=1$. Here $N_\rmI\geq 0$ is a free parameter, which can be varied. 

Expectation values are obtained by averaging over the noise; they
evolve according to
\be
\bra O\ket_{P(t)} = \int dxdy\, P(x,y;t)O(x+iy),
\ee
where the distribution $P(x,y;t)$ satisfies the FPE
\be
\label{eq:FP}
\dot P(x,y;t) = L^TP(x,y;t),
\ee
with the FP operator ($K_x = -\re\partial_z S(z)$ and 
$K_y = -\im\partial_z S(z)$ are the drift terms):
\be
\label{eq:FPop}
L^T =  \partial_x
\left( N_\rmR\partial_x-K_x\right) +
\partial_y
\left( N_\rmI\partial_y-K_y\right).
\ee

In CL dynamics, convergence to the correct result is not guaranteed. A
necessary condition is that the so-called {\it criteria for correctness} are
satisfied~\cite{arXiv:0912.3360,arXiv:1101.3270}:
\be
C_O \equiv \left\bra\tilde L O(z)\right\ket=0,
\ee
in principle for a complete set of holomorphic observables $O(z)$. Here
$\tilde L$ is the Langevin operator
\be
\tilde L = \left[\partial_z - (\partial_zS(z))\right]\partial_z.
\ee
We consider as observables $O_n(z) = \frac{1}{n}z^n$, with $n$ even. 

\begin{figure}[t]
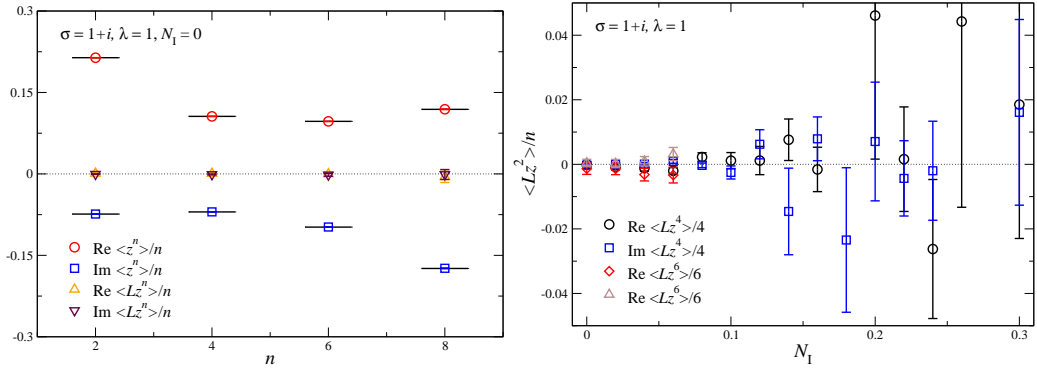

\begin{center}
\epsfig{figure=./figures/plot-CC-A1B1l1NI0-v2.eps,width=0.44\textwidth}
\epsfig{figure=./figures/plot-CC-A1B1l1vsNI-v5.eps,width=0.46\textwidth}
\vspace{-2mm}
\end{center}
 \caption{(Left) Real and imaginary parts of the expectation values 
$\frac{1}{n}\bra z^n\ket$ and criteria for correctness  
$C_n=\frac{1}{n}\bra \tilde Lz^n\ket$ versus $n$ at $\sigma=1+i$ and 
$\lambda=1$ for real noise ($N_\rmI=0$).  The horizontal lines 
indicate the exact value. (Right) Criteria for correctness as a 
function of $N_\rmI$ for $n=4,6$.}
 \label{fig:CC}
\end{figure}

In Fig.~\ref{fig:CC} (Left) CL results are shown for the real and imaginary 
parts of the observables $\frac{1}{n}\bra z^n\ket$ and for the  criteria 
for correctness  $C_n=\frac{1}{n}\bra \tilde Lz^n\ket$, for $n=2,4,6,8$. 
The figure shows the result for real noise, $N_\rmI=0$, and parameters
$\sigma=1+i$ and $\lambda=1$: all expectation 
values agree with the exact result, denoted with the horizontal lines, 
and the criteria for correctness are all consistent with zero, 
as it should be. 
Moreover, we have studied how the observables and the criteria for 
correctness depend on the amount of complex noise. 
For small $N_\rmI$ the observables with $n=2,4$ appear to be consistent with 
the exact result, while for larger $N_\rmI$ they start to deviate. 
Problems can be detected by considering higher moments.
For small $N_\rmI$ the observables (with $n\geq 6$) and the criteria 
(with $n\geq 4$) are only marginally consistent with the expected results, 
while for larger $N_\rmI$ they suffer from large fluctuations, 
see Fig.~\ref{fig:CC} (Right), and can no 
longer be sensibly determined. According to the analytical 
justification~\cite{arXiv:0912.3360,arXiv:1101.3270}, this implies that the 
results from CL cannot be trusted in the presence of complex noise.

\section{Probability distributions}

In this section we study the distribution $P(x,y)$ in the complexified space.
It can be obtained directly collecting histograms of the
distribution during the CL evolution; this method can be easily extended 
to multi-dimensional integrals and field theories. 
A different method to determine $P(x,y)$ was followed in 
Ref.~\cite{Duncan:2012tc}: the idea is to expand the distribution in terms 
of a truncated set of basis functions and solve the resulting matrix problem 
numerically.
If we denote the eigenvalues of $-L^T$ with $\kappa$ and the 
eigenfunctions with $P_\kappa(x,y)$, what we have to do is to 
solve the following 
eigenvalue problem: $-L^T P_\kappa(x,y) = \kappa P_\kappa(x,y)$.
The time-dependent distribution can then be written as:
\be
P(x,y;t) = P_0(x,y) + \sum_{\kappa\neq 0}  e^{-\kappa t}  P_\kappa(x,y),
\ee
where  $P_0$ is the ground state, i.e. the one with eigenvalue $\kappa=0$. 
Note that, in order for $P_0(x,y)$ to be the equilibrium distribution, it
is necessary that for all other eigenvalues $\re \kappa>0$.

Following Ref.~\cite{Duncan:2012tc}, we expand $P(x,y)$ in a basis 
of Hermite functions, i.e. 
\be
P(x,y) = \sum_{k=0}^{N_H-1}\sum_{l=0}^{N_H-1} c_{kl} H_k\left(\sqrt{\omega}x
\right) H_l\left(\sqrt{\omega}y\right),
\label{eq:distr}
\ee
where $\om$ is a variational parameter appearing in the harmonic oscillator 
eigenfunctions, and $N_H$ indicates the number of Hermite functions included 
in the truncated basis. The coefficients $c_{kl}$ have to be determined;
this is explained in detail in Ref.~\cite{Aarts:2013qia}.

\begin{figure}[t]
\begin{center}
\epsfig{figure=./figures/plot-kappa-A1B1l1-NI1-NH150.eps,width=0.46\textwidth}
\epsfig{figure=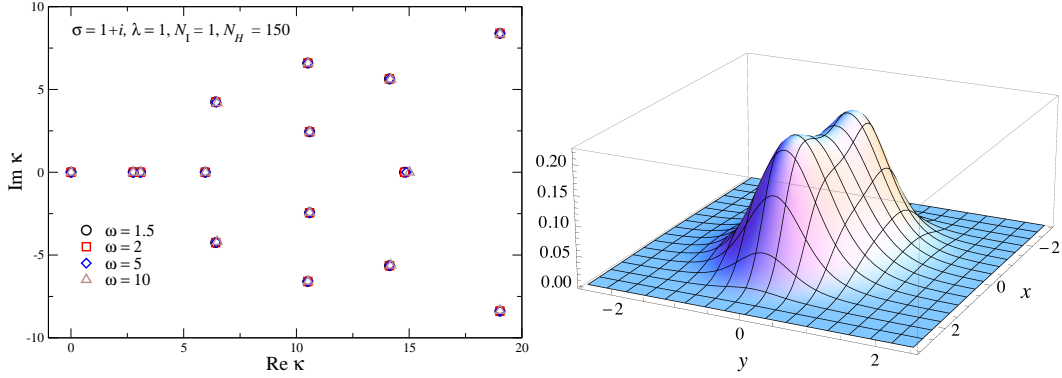,width=0.46\textwidth}
\vspace{-2mm}
\end{center}
 \caption{
(Left) Eigenvalues of the FP operator $-L^T$ for complex noise, with $N_\rmI=1$,
magnified around the smallest eigenvalues, for various values of $\om$, at 
$\sigma=1+i$, $\lambda=1$, and $N_H=150$. (Right) Distribution $P(x,y)$ in 
the $xy$-plane for complex noise, with $N_\rmI=1$ and $\om=1.5$.}
 \label{fig:kappa-NI1}
\end{figure}

\begin{figure}[t]
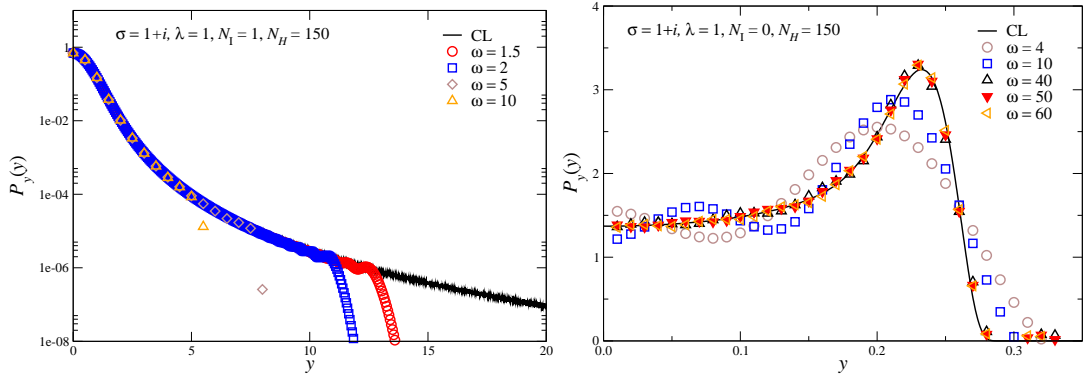

\begin{center}
\epsfig{figure=./figures/plot-Py-A1B1l1-NI1-NH150.eps ,width=0.48\textwidth}
\epsfig{figure=./figures/plot-Py-A1B1l1-NI0-NH150-v2.eps,width=0.46\textwidth}
\vspace{-2mm}
\end{center}
 \caption{
Partially integrated distributions $P_y(y)$ for different values of 
$\om$ with complex noise, $N_\rmI=1$ (left plot) and real noise, 
$N_\rmI=0$ (right plot).
In both cases the noisy (black) data was obtained by a CL simulation.}
\label{fig:P-NI1}
\end{figure}

We start with the case of complex noise, $N_\rmI=1$. 
The parameters in the action are 
taken as $\sigma=1+i$ and $\lambda=1$, and we consider a basis with 
$30\leq N_H\leq 150$ Hermite functions and different values of $\om$.
In the limit of large $N_H$ the results are expected to be independent 
of the value of $\om$. In practice however, we find that for finite $N_H$ 
the parameter $\om$ plays the role of a tuning parameter:
in particular, when $\om$ is too small, there are eigenvalues with a 
negative real part.
We find that there is always an $\om$ interval for which:
i) there is an eigenvalue consistent with 0;
ii) the other eigenvalues are in the right half-plane;
iii) the reconstructed ground state distribution is stable under 
variation of $N_H$ and  $\om$. 
The smallest 15 eigenvalues are shown in Fig.~\ref{fig:kappa-NI1} (Left):
for the $\om$ values shown here, all eigenvalues are in the right half-plane 
and the spectrum around the origin is to a good extent independent of $\om$.
The reconstructed distribution $P(x,y)$, using Eq.~(\ref{eq:distr}), is shown 
in Fig.~\ref{fig:kappa-NI1} (Right). We find a smooth distribution with a 
double peak structure.
In Fig.~\ref{fig:P-NI1} (Left) we present the partially integrated 
distributions $P_y(y)=\int_{-\infty}^\infty dx\, P(x,y)$ on a logarithmic scale. 
Besides presenting results for various $\om$ values, we also show  
the histogram obtained during a CL simulation.
We observe an acceptable agreement between the CL results and the solution 
of the FPE for $\om\sim 1.5, 2$, down to a relative size of $10^{-6}$, after 
which the FP solution can no longer cope. We interpret this as a 
manifestation of the truncation. When $\om$ is taken too large, the 
disagreement occurs for smaller values of $y$. 
Moreover, we have verified that both partially integrated 
distributions $P_x(y)$ and $P_y(y)$ are characterised by a power decay
with power 5. 
In the case of real noise, $N_\rmI=0$, we note that in all cases there is 
an eigenvalue at (or close to) the origin, but in general convergence is 
much harder to establish from a study of the eigenvalues alone.
Convergence of $P_y$ as $\om$ is increased is demonstrated in 
Fig.~\ref{fig:P-NI1} (Right) and we observe that a large value of $\om$ is 
required, $\om\sim 50$. It is of course expected that the chosen value of 
$\om$ eventually becomes irrelevant, but for finite $N_H$ keeping 
$\om$ as a tuning parameter is essential. The distribution is very well 
localised and appears to drop to 0 around $y=0.28$.
The distribution $P_x(x)$, in contrast to the case of complex noise, is now
characterised by an exponential rather than a power decay.

We conclude that for this choice of parameters ($\sigma=1+i$ and $\lambda=1$) 
the decay in the case of real noise is manifestly different compared to 
complex noise. In the latter we found a power decay, resulting in ill-defined 
moments $\bra z^n\ket$ when $n>4$, while here we find exponential decay in 
the $x$ direction and, as we will see below,  in the $y$ direction 
support only inside a strip. As a result there is no problem in 
computing higher moments, since they are all well-defined.

\section{Interpretation}

\begin{figure}[t]
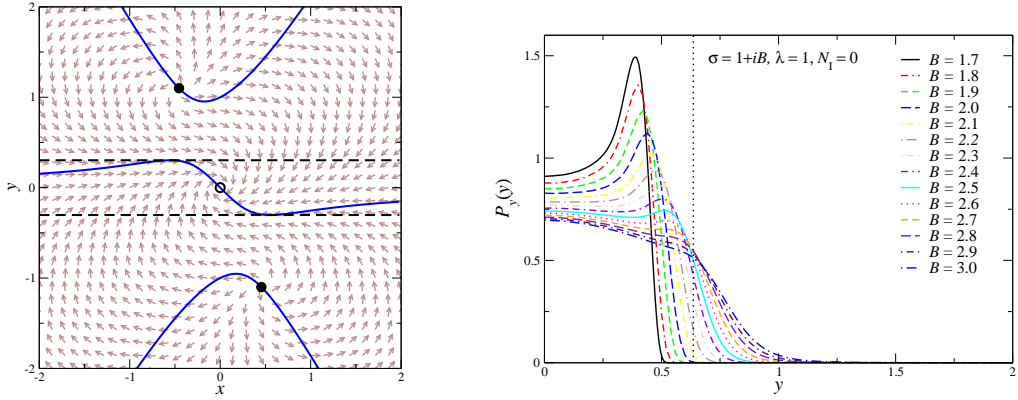

\begin{center}
\epsfig{figure=./figures/plot-flow-A1B1l1-xmgrace4.eps,width=0.35\textwidth}
\hspace{10mm}
\epsfig{figure=./figures/plot-Py-A1l1NI0-vsB-v2.eps, width=0.46\textwidth}
\vspace{-2mm}
\end{center}
 \caption{(Left) Classical flow in the $xy$-plane, for $\sigma=1+i$ and 
$\lambda=1$. The attractive/repulsive fixed points are indicated with 
the open/filled circles. The full lines indicate where $K_y(x,y)=0$. 
The horizontal dashed lines indicate the strip in which the CL process 
takes place in the case of real noise. (Right) 
Distribution $P_y(y)$ for different values of $B$, with $N_\rmI=0$, 
at $\sigma=1+iB$ and $\lambda=1$, obtained with CL;
the vertical line indicates the boundary of the strip for $B=1.7$. 
For larger $B$ values, there is no longer a boundary.}
\label{fig:flow}
\end{figure}

The classical flow diagram is shown in Fig.~\ref{fig:flow} (Left), 
for $\sigma=1+i$ 
and $\lambda=1$. We show the direction of the classical force by an arrow 
pointing in the direction $(K_x(x,y), K_y(x,y))$. The arrows are normalised 
to have the same length. 
There are three fixed points, where $K_x=K_y=0$: an attractive point 
at the origin and two repulsive fixed points.
The flow is directed towards the origin, provided that $|y|$ is not too large.
This can be made more precise by studying where $K_y(x,y)$ changes sign:
this is indicated in the classical flow diagram with full (blue) lines.
We now realise that along the horizontal dashed lines, which are determined 
by the extrema of the centre curve where $K_y=0$ ($y=\pm0.3029$ in this case), 
the flow is always pointing inwards, i.e. towards the real axis. 
In absence of a noise component in the vertical direction, this creates a 
barrier for the Langevin evolution beyond which it cannot drift.
We conclude therefore that in the case of real noise the process takes 
place in the strip determined by $-0.3029<y<0.3029$.
This is consistent with the conclusions drawn above from the histograms 
and the FPE solution of the distribution $P(x,y)$.
In the presence of complex noise, this conclusion no longer holds and the 
entire $xy$-plane can be explored.

As shown in Ref.~\cite{Aarts:2013qia} it is possible to make the argument 
based on classical flow presented above rigorous and show directly from the 
FPE that the equilibrium distribution $P(x,y)$ is strictly zero in strips in 
the $xy$-plane. To summarise, for real noise, we find the following:
\begin{enumerate}
\item when $3A^2>B^2$,  $P(x,y)=0$ for $y^2 > \frac{A}{2\lambda}
\left( 1 - \sqrt{1-\frac{B^2}{3A^2}}\right)$;
\item  when $3A^2<B^2$, there are no restrictions on $P(x,y)$. 
\end{enumerate}
Interestingly, the derivation in Ref.~\cite{Aarts:2013qia} demonstrates 
that when $3A^2<B^2$, i.e.
in absence of strips, one may therefore expect a breakdown of CL with real 
noise, similar as with complex noise. This is indeed what happens: as shown in 
Fig.~\ref{fig:flow} (Right) where the distribution $P_y(y)$ is plotted
as $B$ is increased. The delocalisation  has a detrimental 
effect on the results of the CL process. Studying the moments and the 
criteria for correctness in this case, we observe that increasing $B$ 
has a similar effect as increasing $N_\rmI$: moreover a power law takes place
again with power 5. 
In addition in Ref.~\cite{Aarts:2013qia}, we have shown that it is possible 
to understand the universal power decay directly from the FPE: as a consequence,
for large $|x|$ and $|y|$, the distribution decays as a power, according to 
$P(x,y) \sim {(x^2+y^2)^{-3}}$.
We conclude that in absence of strips a universal power law decay is 
present, which results in a breakdown of the formal 
justification~\cite{arXiv:0912.3360,arXiv:1101.3270} and wrong or wildly 
fluctuating results in practice.

\section{Conclusion}
In order to justify the results obtained with complex Langevin dynamics, 
it is necessary that the probability distribution is sufficiently 
localised in the complexified configuration space. Here we have studied 
properties of this distribution via a number of methods, in the case of 
a simple model. 
In this case the FPE can be solved explicitly, via an expansion in a 
truncated set of basis functions. However, it is still a nontrivial 
problem and perhaps the best way to find the distribution is by brute 
force, i.e. during the CL simulation.
We have demonstrated that the essential insight can already be obtained 
from a combination of histograms of partially integrated distributions 
and the criteria for correctness, which gives a consistent picture of the 
dynamics. These tools are readily available in field theory. 

Recently, a new approach, based on deforming the integration contour of 
the path integral into the complex plane and performing Monte Carlo 
simulations on the so-called Lefschetz 
thimbles, has been introduced~\cite{Cristoforetti:2012su,Mukherjee:2013aga}. 
A comparison between CL and the Lefschetz thimble for the model 
considered here, can be found in Ref.~\cite{Aarts:2013fpa}.

Finally, we remark that our conclusions are also immediately applicable 
to nonabelian SU($N$) gauge theories~\cite{Sexty:2013ica}, for which 
gauge cooling~\cite{Aarts:2013uxa,Seiler:2012wz}
provides a means to control the distribution in 
SL($N,\mathbb{C}$), a possibility not present in simpler models.

\end{document}